\documentclass[pra,twocolumn,superscriptaddress,floatfix,amsmath,nofootinbib,amssymb]{revtex4-1}

\usepackage{graphicx}
\usepackage{bm}
\usepackage{verbatim}
\usepackage{mathrsfs}
\usepackage{color}

\newcommand{\ii}{\mathrm{i}}

\newcommand{\ket}[1]{| {#1} \rangle}
\newcommand{\ketno}[1]{\big| {#1} \big\rangle}
\newcommand{\bra}[1]{\left\langle {#1} \right|}

\begin{document}

\title{Mode Invisibility and Single Photon Detection}
\author{Marvellous Onuma-Kalu}
\email{monumaka@uwaterloo.ca}
\affiliation{Department of Physics \& Astronomy, University of Waterloo,  Ontario Canada N2L 3G1}
\author{Robert B. Mann}
\email{rbmann@uwaterloo.ca}
\affiliation{Department of Physics \& Astronomy, University of Waterloo,  Ontario Canada N2L 3G1}
\author{Eduardo Mart\'{i}n-Mart\'{i}nez}
\email{emartinm@uwaterloo.ca}
\affiliation{Institute for Quantum Computing, University of Waterloo, Waterloo, Ontario, N2L 3G1, Canada}
\affiliation{Department of Applied Math, University of Waterloo, Waterloo, Ontario, N2L 3G1, Canada}
\affiliation{Perimeter Institute for Theoretical Physics, 31 Caroline St N, Waterloo, Ontario, N2L 2Y5, Canada}

\begin{abstract}
We propose a technique to probe the quantum state of light in an optical cavity without significantly altering it.  We minimize the interaction of the probe with the field by arranging a setting where the largest contribution to the transition probability is cancelled. We show that we obtain a very good resolution to measure photon population differences between two given Fock states by means of atomic interferometry.
\end{abstract}

\maketitle
\section{Introduction}

Determining the state of a quantum system without significantly perturbing it can indeed pose a challenging problem. One example is the destructive process involved in the detection of light trapped in an optical cavity. Using a photodetector, which in its most schematic version could be just a single atom, if we want to probe a quantum state containing a handful of photons, it is very likely that in order for the detector to click, photons have to be absorbed from the field by the atom and therefore the state of the light inside the cavity would be drastically altered if the number of photons is small enough.

These limitations have been overcome with the proposal of quantum non-demolition schemes (QND), with which it is possible to carry out weak measurements that obtain the most information possible from a quantum state without significantly altering the state \cite{Braginski,firstdemo,secondemo}. 
In these schemes, a QND measurement was adapted to the counting of light quanta. A beam of Rydberg atoms evolves into a superposition of two distinct states and is then made to interact with light sustained in a highly reflective cavity and non-resonant with the energy gap between the two atomic states. In this way the field in the cavity remains intact while the atoms successfully cross the cavity one by one. The  fact that the atomic gap is non-resonant with the field mode to be probed  ensures a nondemolition measurement on the trapped photons but also yields an observable shift in atomic energy levels proportional to the photon number that is measurable using a Ramsey atomic interferometer \cite{Haroche, Sergelecture}. Indeed, the use of atomic interferometry also promises impressive advances in metrology, from the measurement of space-time curvature of the Earth (see for instance \cite{Rideout2012} for a quick review)  to  proposals for improving  the detectability of the Unruh effect by reducing required accelerations, 
to the implementation of a highly sensitive quantum thermometer \cite{Quantumthermometer}.

Inspired by these results we propose to use a similar scheme to that of \cite{Unruh,Quantumthermometer} for the detection of photons. We will not use geometrical phases but dynamical phases, and we will not make any of the common idealizations on the quantum field behaviour (i.e. we will not carry out single-mode nor rotating wave approximations \cite{Robert2013,ScullyBook}). We will introduce a technique, which we will call `mode invisibility' to minimize the backreaction of an atom on a state of the electromagnetic field while maximizing the information about the field state that we can extract by observing the atom. For this we will use atoms that are not highly excited but instead in a relatively stable ground state, which will acquire a global phase after interacting with the field that shall be measured a simple interferometric scheme.

 The method pursues the same objective as the celebrated work by Serge Haroche \cite{Haroche,cavities,nature,Sergelecture}: to get a non-negligible phase without altering the state of the field. In the methods introduced by Haroche, this is accomplished by using atoms that are largely off-resonant with respect to the field mode that we want to probe. The main difference with the technique presented in this paper is that in our case we minimize the back reaction of the atom in the field while keeping the atomic probes on resonance with the mode we want to probe. As we will discuss below, being on resonance with the probed mode   helps increase the strength of the phase acquired by the atomic probe.  

In this article, we propose a new atomic interferometry method aimed to determine the difference between the photon content of two given Fock states, allowing therefore to find the photon content of a Fock state of light contained in a cavity by comparison with a known state. We would like to be able to distinguish states of light containing very few photons without  significantly altering the photon content of the field. We will present techniques where the probability of atomic probes altering the state of light during their interaction with it in the cavity could be minimized even while maximizing the global phase change carried by the probe.

\section{The setting}\label{sectionmeth}

\begin{figure}[h!]
\begin{center}
\includegraphics[width=.50\textwidth]{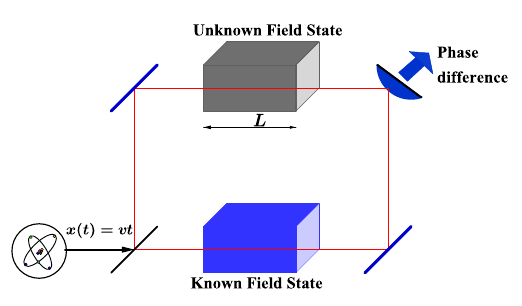}
\caption{(Color online): Schematic set-up of our quantum system interaction. An atom initially in its ground state $|g\rangle$ is   made to propagate along the two branches of an atomic interferometer via a beamsplitter. In the two branches, each partial state encounters a cavity of length $L$ containing multiple quantum field modes. Different phases are acquired along each arm due to interaction with the quantum field. The phase difference measured at the output of the interferometer determines the unknown field state \cite{Quantumthermometer}}
\label{atominterferometer}
\end{center}
\end{figure}

 Fig. \ref{atominterferometer} shows the set-up for our scheme. Two optical cavities each of length $L$, containing   known and   unknown quantum field states respectively, are placed along the two branches of an atomic interferometer. We send single atoms initially in their ground state into this atomic interferometer through a beamsplitter so that they propagate along the two branches of the atomic interferometer, respectively interacting with the known and unknown field states placed along their paths. After an interaction time $T$ the atom exits the cavities, the interaction between the field modes and the single atom creating a phase change in the atom's quantum state. The atomic phase change acquired in the two optical cavities differs. By looking at the interferometer's output at the point where the partial beams recombine, the measured phase difference can be expected to reveal the unknown field state. This  kind of set-up has been considered previously as a way to detect the Unruh effect  \cite{Unruh} and to measure, with great precision, the temperature of a hot source relative to some reference source \cite{Quantumthermometer}. In either case, the phase acquired by the detector differs for different quantum field states. For instance, in the case of a thermal field state, this phase encodes information about the field's temperature. The sensitivity to temperature is very strong, and it has been shown that measurement of the phase can in principle be used  to detect the Unruh effect at accelerations nine orders of magnitude lower than previous proposals.
 
We propose to use this phase difference to measure the photon number of an unknown bosonic field state. Regarding the atom's quantum state as a wave, the peaks and the dips  of the wave become shifted during interaction with the discrete quantum field modes. As we shall demonstrate,  the phase difference can be significant and thus measured.  We map the field state (namely, the photon content of the state) in the cavity onto the global phase difference acquired when the atom exits the cavity (see Fig. \ref{atominterferometer}).  The phase $\gamma$ is then a function of the  photon number $n$.  If the state of the field is not significantly altered, then repeated measurement can be used to estimate the phase,  thereby revealing the exact number of photons in the unknown field state \cite{nature}. 

In our measurement scheme, we consider the joint atom-quantum field state at time $t=0$  to be $|\psi(0) \rangle = |g, n_{\alpha}\rangle$ where we set the field state $|n_{\alpha}\rangle$  (with $\alpha = 1,2, \cdots \infty$) as the eigenstate of our system's  photon number observable, corresponding to $n$ photons in the cavity mode $\alpha$ of frequency $\omega_\alpha=\alpha\pi/L$.
We assume that single atoms enter the cavity in their ground state $|g\rangle$. By setting the cavity mode frequency $\omega_{\alpha}$ at resonance with the atom transition frequency $\Omega$, the joint system therefore undergoes oscillations at angular frequencies $(\omega_{\alpha} \pm \Omega)$ between various possible states.

The heart of our work lies in our ability to ``manipulate''  the interaction between these single atoms and field modes trapped in the optical cavity without perturbing the combined quantum system very much. 
These systems are assumed to be coupled by an interaction Hamiltonian for a short time $T$ by a unitary evolution operator $U(0,T)$, so that the joint atom-field state after the interaction is $|\psi(T)\rangle = U(0,T)|\psi (0)\rangle$.  We require that the coupling to the cavity modes is taken to be weak enough so that the effect of the atom-field interaction when the atom flies through the cavity for short times does not alter the probability distribution of the joint quantum state.

 During this interaction, it is known that the atom acquires a phase \cite{Unruh} due to the presence of the quantum field. 
We will calculate the phase acquired by the atom for a given field mode, check that the phase difference obtained for different field modes is measurable, and see if this difference responds to changes in the number of field modes. To proceed, we consider the interaction Hamiltonian, $H_{I} = \lambda  \mu(t) \phi[x(t)] $, which describes a point-like two-level quantum system with monopole moment  $\mu(t) = (\hat{\sigma}_{+} + \hat{\sigma}_{-})$ coupled to a scalar field  $\phi[x(t)]$ along the detector's trajectory  $x(t)$ with coupling strength $\lambda$  \cite{Unruh, Quantumthermometer,AasenPRL}. This light-matter interaction model, known as an Unruh-DeWitt detector, has been shown to model an atom coupled to the electromagnetic field as long as there is no angular momentum exchange involved \cite{Wavepackets}.

\section{Phase and transition probabilities}

In our setting, $x(t)$ represents the atom's trajectory inside a one-dimensional perfectly reflective cavity with boundary at $x = 0$ and $x = L$. In the interaction picture, the Hamiltonian describing the atom-field interaction can  be written as:
\begin{align}\label{interactionhamiltonian}
H_{I} = &  \lambda  \mu(t) \sum_{\beta=1}^{\infty}( a_{\beta}^{\dagger} e^{\ii\omega_{\beta} t} + a_\beta e^{-\ii\omega_{\beta} t})  \frac{\sin[k_{\beta}x(t)]}{\sqrt{k_\beta L}},
\end{align}
where the field is expanded in terms of the stationary wave modes of the Dirichlet cavity. The time evolution under this interaction Hamiltonian from a time $t=0$ to a time $t=T$ is given by
\begin{align}\label{unitary}
U(T,0) = \mathcal{T}\text{exp}\bigg[\frac{1}{\ii}\int_{0}^{T} dt H_{I}(t)\bigg].
\end{align}
After a time $T$, the state of the system  is given as $\ket{\psi(T)} = U(0,T)\ket{\psi(0)}$. Perturbatively expanding the time-ordered exponential  we obtain  $\ket{\psi(T)} = \ket{\psi(0)}+\ketno{\psi^{(1)}_T}+\ketno{\psi^{(2)}_T} +\mathcal{O}(\lambda^3)$ where
\begin{align}
\ketno{\psi^{(n)}}= U^{(n)}\ket{\psi(0)},
\label{four}
\end{align}
with the different order contributions to $U(0,T)$ being
\begin{equation}\label{eq:pert}
U(0,T) \!=\! \openone\!\underbrace{-\ii\!\!\int_{0}^{T}\!\!\!\!\!d t_1 H_{I}(t_1)}_{U^{(1)}}\underbrace{ -\!\!\int_{0}^{T}\!\!\!\!\!dt_1 \!\!\int_{0}^{t_1}\!\!\!\!\!dt_2\,    H_{I}(t_1) H_{I}(t_{2})}_{U^{(2)}}+\hdots
\end{equation}


The field-atom system starts from an initial state $\ket{\psi(0)}$ and  traverses  the cavity during a time $T$. Requiring the probability that the whole system remains in the same state to be approximately unity, i.e,
\begin{align}\label{hypo}
\big | \langle \psi (0) | U(0,T)| \psi(0) \rangle \big | ^{2} \approx 1,
\end{align}
ensures   the minimal possible alteration to the field state.  
Under this assumption the final state of the system would be very approximately equal to the initial state except for a global dynamical phase
\begin{align}\label{phased}
|\psi(T) \rangle = U(0,T)|\psi(0) \rangle \approx e^{i \gamma}|\psi(0) \rangle,
\end{align}
where $\gamma$ is the phase factor to be determined. In particular the state of the measuring device (the single atom) remains the same before and after it exits the cavity except for a dynamical phase. Even in regimes where the interaction between a single atom and the quantum field is so weak such that the field state to be measured is unperturbed, we expect that the measured phase value will hold information about this field state.

For this scheme to work we have to make sure that (a) the hypothesis \eqref{hypo} holds and (b)  the phase is significantly measurable in the regimes where this is so. Even with these conditions satisfied, we note that the information that can be extracted from the joint system is limited: the only way to measure a global phase is by means of an interferometry experiment. To this end we have to compare the state of the field with another known state. As the phase is a scalar magnitude, we would be able to partially identify states within a known 1-parametric family of states. By so doing, the measured phase after the interaction will indeed be higher.  Nevertheless we will see below that these phases are much more sensitive to the state of the field than are the probabilities of transition.


We want to determine the number of photons with great precision, perturbing the system as little as possible. An ideal projective measurement over sectors of $n$ particles is physically unfeasible. In practice this would involve detectors coupled to the field. By the time a field quanta is detected (for  example by counting a detector click), the back reaction of the detector would have modified the state of the field, which will no longer be in a Fock state but instead in a mixed superposition of different quanta sectors (see for instance \cite{Robert2013}). 

Let us assume that the state of the field is prepared in a cavity in an unknown Fock state of $n$ photons in the cavity mode $\alpha$ of frequency $\omega_\alpha=\alpha\pi/L$. All the rest of the modes are prepared in the vacuum (or in very low-populated states). Given that our detector is prepared in the ground state, we have
\begin{align}\label{initialstate}
\ket{\psi(0)}=\ket{g}\otimes \ket{n_\alpha}\bigotimes_{\beta\neq \alpha}\ket{0_\beta}.
\end{align}

The first step is to estimate the probability of an atomic transition while crossing the cavity. We want this to be approximately zero. For that we calculate the different contributions to the leading order expansion of the time evolution operator.  The first order contribution to the evolution operator \eqref{unitary} is
\begin{align}
\nonumber U^{{(1)}} = \sum_\beta \frac{\lambda}{\ii} &\Big( \sigma^{+} a_{\beta}^{\dagger}X_{+,\beta} + \sigma^{-} a_{\beta}X_{+,\beta}^{*}+\sigma^{-} a_{\beta}^{\dagger}X_{-,\beta}\\
&+\sigma^{+} a_{\beta}X_{-,\beta}^{*} \Big),
\end{align}
where for notational convenience we have
\begin{align}\label{prob}
X_{\pm, \beta} =  \frac{1}{\sqrt{k_{\beta}L}}\int_{0}^{T} dt e^{\ii(\pm \Omega + \omega_{\beta})t}\sin[k_{\beta}x(t)].
\end{align}
The argument in the exponential describes rotating  \text{$(\omega_{\alpha} - \Omega)$} and counter-rotating  \text{$(\omega_{\alpha} + \Omega)$}  terms, with the typical resonance condition $\omega_{\alpha} = \Omega$. 

After some lengthy but straightforward algebraic calculation (detailed in appendix \eqref{anything}) we can find that the probability of a transition to an excited state is 
\begin{align}\label{transitionp}
P_{\ket{e}}(T)\!=\!{\lambda^{2}}\Bigg[|X_{-,\alpha}|^{2}n+ |X_{+,\alpha}|^{2}(n+1)\! +\! \sum_{\beta\neq \alpha} |X_{+, \beta}|^{2} \Bigg]
\end{align}
The three contributions to this probability can be roughly understood as follows. The first term corresponds to the contribution from the excitation probability due to  the detector absorbing a photon from the field mode $\alpha$. The second corresponds to the atom getting excited and emitting a photon to the mode $\alpha$. This is the typical counter-rotating contribution. The third term corresponds to the vacuum fluctuations due to the rest of the modes (see for instance \cite{ScullyBook, Robert2013}). Assuming the detector is tuned to be resonant with the mode of the field we want to probe, the largest contribution would come from the first term.  This is the principal contribution that can jeopardize the hypothesis (see equation \eqref{hypo}). However  we will propose a technique by which we can identically cancel this contribution if the unknown state (Fig. \ref{atominterferometer}) is prepared in an even harmonic of the cavity  ( as we shall show later in section \ref{stealth}) so that $P_{\ket{e}}(T)\ll 1$.

If the probability of transition is approximately zero, the first contribution to the phase would come from $|\psi^{(2)}_T\rangle$, namely from the second order terms that are proportional to the initial state. In detail, the second order correction is given by
\begin{align}\label{crossterms}
\nonumber|\psi_T^{(2)}\rangle&\!= \!-\lambda^{2} \Big[ n\frac{C_{-,\alpha}}{k_{\alpha}L}+\! \sum_{\beta\neq \alpha} \frac{C_{+, \beta}^{*}}{k_{\beta}L} +  (n+1)\frac{C_{+,\alpha}^{*}}{k_{\alpha}L}\Big]\ket{\psi(0)}\\
&+\ket{\psi(T)}_\perp,
\end{align}
where
\begin{align*}
C_{\pm, 
\beta} = & \int\limits_{0}^{T} \mathrm{dt}\int\limits_{0}^{t}\mathrm{dt'}~ e^{\ii(\omega_{\beta} \pm \Omega)(t-t')}\sin[k_{\beta}x(t)]\sin[k_{\beta}x(t')]
\end{align*}
and where  $\ket{\psi(T)}_\perp$  is the second order contribution that  is orthogonal to the initial state and which is irrelevant to the computation of the phase.  
Of course the part that we are not specifying, i.e. $\ket{\psi(T)}_\perp$, should be small enough for all our assumptions to hold. Its magnitude will have an impact on the visibility of the fringes in the interferometric experiment as we discuss below. We will see that we will be able to compute the effect of this term and keep it under control.

\section{Mode invisibility}\label{stealth}

For our proposed  experiment to be feasible we must maximize the phase acquired by the atom flying through the cavity. This seems to suggest that we must prepare our detector such that it is resonant with the mode of the field we would like to probe.  However since the resonant mode gives the largest contribution to the detector's transition rate, this also gives in general  a higher probability of altering the state of the cavity,  jeopardizing the approximation made in order to apply our formalism. 

\begin{figure}
\begin{center}
\includegraphics[width=.25\textwidth]{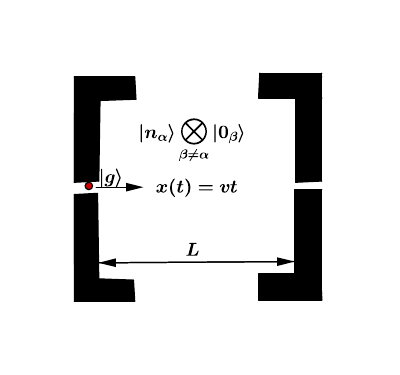}\includegraphics[width=.25\textwidth]{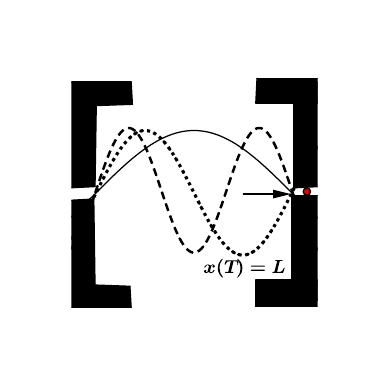}
\caption{(Color online): Scheme of an atom in its ground state $|g\rangle$ going through a cavity at constant speed. The cavity modes are in a state $ \ket{n_\alpha}\bigotimes_{\beta\neq \alpha}\ket{0_\beta}$.  Atoms resonant with even modes preserve our approximation \eqref{phased} and   acquire a significant global phase.}
\label{waves}
\end{center}
\end{figure}

Fortunately we can use the spatial distribution of the field modes to our advantage in order to minimize the effect of the resonant mode on the transition probability and still have a strong contribution to the phase.The idea is to take advantage of the spatial symmetry of field modes so that the  atoms interact with light in a non-destructive way \cite{Dicke55,Stephen98, Sarkisyan04}.  
When the atom interacts with an even mode of the cavity (like the second harmonic showed in the dotted line of Fig \ref{waves}), most of the changes that it will introduce in the field state while flying half the way through the cavity $x\in [0, L/2]$ will be undone when the atom flies through the second half $x\in [L/2, L]$. As a first approximate description of the phenomenon, whatever the atom absorbs  while flying through the first half of the cavity will be identically re-emitted while flying through the second half so that the state of field and atom are the same modulo a phase. This is possible because the effective sign of the coupling to the cavity ($\lambda$ times the spatial distribution of the mode) reverses half way through the flight path of the atom. 

Of course  in our particular setting this will only be true for the first order terms of the perturbative expansion, since the even orders in the coupling strength $\lambda$ will not see this effective sign change. This will have the advantage that we can cancel out the leading order contribution to the transition amplitude for the field and the detector while keeping constant the leading order in the phase effects. 

Let us see the points made above in a rigorous approach: to leading order in perturbation theory the transition probability is given by the expression \eqref{transitionp}. If  we send an atom through a cavity at a constant speed $v$, the time it spends inside the cavity is $T=L/v$. 

If we introduce this into equation \eqref{prob}, also taking into account that  $k_\beta=\beta\pi/L$, we obtain
\begin{align}\label{inv1}
X_{\pm, \beta} =  \frac{1}{\sqrt{\pi {\beta}}}\int_{0}^{L/v} \text{d}t\, e^{\ii(\pm \Omega + \omega_{\beta})t}\sin\left(\frac{\beta\, \pi}{L}vt\right).
\end{align}
The integral can be readily solved, giving
\begin{align}\label{inv2}
X_{\pm, \beta} =  \frac{\left[e^{\ii\frac{L}{v}(\omega_\beta\pm\Omega )}(-1)^{\beta} -1\right] L v\sqrt{\beta\, \pi} }{\left(\beta\, \pi v\right)^2-L^2(\omega_\beta\pm\Omega)^2}
\end{align}

As  is well-known \cite{ScullyBook}, the largest contribution to the transition probability comes from the rotating-wave term $X_{-, \beta}$ for the resonant mode, i.e. $\omega_\beta=\Omega$, where the frequency of the mode coincides with the atomic gap. Indeed, as seen in \eqref{inv2}, all the counter-rotating contributions $X_{+, \beta}$  for all $\beta$ and the off-resonant rotating-wave contributions  $X_{-, \beta} $ for all $\omega_\beta\neq\Omega$ are damped by the square of the sum or the difference between frequency of the atomic gap and the frequency of the cavity modes. Therefore the transition probability would be damped by a quartic factor on the sum of difference of frequencies. 

That would be a problem: if we want to maximize the phase we better prepare the state to probe in a resonant mode of the cavity, but that will involve increasing the probability of transition and therefore taking us out of the adiabatic approximation discussed in section \ref{sectionmeth}.

To keep the probability of transition the smallest possible we can use the spatial distribution of the modes to our advantage: the multiples of the second harmonic  ($\beta=2,4,\dots$) have the property that their spatial function is of odd parity, evident in the sine factor in \eqref{interactionhamiltonian}. Therefore an atom that flies through the cavity would be effectively changing the sign of the coupling by an even number of times as   illustrated in Fig. \ref{waves}.

This is seen in expression \eqref{inv2}. For example, for $X_{-, \alpha}$ (corresponding to the resonance: $\omega_\alpha=\Omega$)  we get
\begin{align}\label{transprob2}
X_{-, \alpha} =  \frac{\left[(-1)^{\alpha} -1\right] L }{\left(\alpha\, \pi \right)^{3/2} v}.
\end{align}

If the field mode  resonant with the detector is an odd harmonic of the cavity (i.e. $\alpha=1,3,\dots$) then this would indeed be the largest contribution to the transition probability. However, if the resonant mode  is an even harmonic (i.e. $\alpha=2,4,\dots$) then the contribution becomes identically zero: $X_{-, \alpha}=0$.  Hence provided   the highly excited state we wish  to probe is prepared in one of those even modes, the mode is invisible to the atom (at leading order in perturbation theory)   and therefore will not perturb it. 

 The  mode invisibility technique is robust against a slight detuning from resonance (i.e. if  $\omega_{\alpha} -\Omega = \delta$). From \eqref{inv2} it is easy to see that the largest contribution to the amplitude of  transition probability  for small $\delta$  becomes
\begin{align}
X_{-, \beta} \simeq  \frac{\ii L^2 \delta \sqrt{2\pi n} }{\left( 2\pi n v\right)^2-L^2\delta^2}\simeq \frac{\ii L^2 \delta }{v^2\sqrt{(2\pi n)^3 }}+\mathcal{O}(\delta^2),
\end{align}
when $\beta=2n$. Since the deviation in the contribution of this term to the probability of transition ($\left|X_{-, \beta}\right|^2$ ) is proportional to   $\delta^2$,  the mode invisibility transition probability cancellation  holds if we are slightly off-resonance.

 We call this method `mode invisibility' because the even mode that is probed becomes effectively invisible to a single detector crossing the cavity completely: its  transition amplitude will not be noticeably altered after crossing the cavity. This single detector will not be excited by the probed mode; instead, it will acquire a global phase that is completely undetectable unless some sort of interferometric experiment involving two atoms is set up. 

 We can improve these results by controlling the speed at which the atom crosses the cavity. If  {$\alpha=2j$} is an even number and the atomic speed is close to an integer divisor of the speed of light, or more concretely if the speed is approximately  $v= 2j c/N$ where $N = 1,2,3 \cdots$ (see  \eqref{gerome}) we can cancel the contribution of $X_{+,\alpha}$ to the transition probability, eliminating any dependence it has on the probed mode. This makes the  mode completely `invisible' to the detector.


\section{Evaluating the phase factor}\label{phasefactors}

Provided the assumptions ensuring \eqref{hypo} are satisfied,  our quantum system evolves after a time $T$ to  a general state of the form
 \begin{align}\label{globalphase}
| \psi (T)\rangle =\operatorname{U(0,T)}|\psi(0)\rangle \approx e^{\ii\eta}|\psi (0) \rangle,
\end{align}
where $\eta$ will be given by
\begin{align}\label{phasecal}
\eta &= -i \operatorname{Ln} \langle \psi(0)|\operatorname{U(0,T)}|\psi(0)\rangle \\
\!\!\!&=\! -i \operatorname{Ln} \bigg( 1\! -\! {\lambda^{2}} \bigg[n\frac{C_{-,\alpha}}{k_{\alpha}L}+\! \sum_{\beta\neq \alpha} \frac{C_{+, \beta}^{*}}{k_{\beta}L} +  (n+1)\frac{C_{+,\alpha}^{*}}{k_{\alpha}L} \bigg]\bigg). \nonumber
\end{align}
Note that $\eta$ is not a real number (so strictly speaking, it is not technically a phase) because the second order correction has a contribution orthogonal to the initial state, as seen in \eqref{crossterms}. The phase factor $\gamma$ to be determined is given by 
\[\gamma=\text{Re}{(\eta)}\]
We also have a contribution $\exp[-|\text{Im}(\eta)|] $ in the component of $| \psi (T)\rangle $ proportional to $|\psi (0) \rangle$. In practical terms  translates into a loss of visibility in the interference pattern.   The desirable regime is therefore   $\text{Im}(\eta)\ll 1$. 

To obtain the leading-order approximation for the phase, let us first assume that we work in a regime where \mbox{$|\eta|\ll1$}, which is consistent with our approximations. If that is the case we can expand the exponential term in \eqref{globalphase}, yielding
\begin{equation}\label{taylor}e^{i\eta} \sim 1+\ii\eta= 1 - \! \lambda^{2} \bigg[n\frac{C_{-,\alpha}}{k_{\alpha}L}+\! \sum_{\beta\neq \alpha} \frac{C_{+, \beta}^{*}}{k_{\beta}L} +  (n+1)\frac{C_{+,\alpha}^{*}}{k_{\alpha}L} \bigg]\end{equation}
and so the phase $\gamma=\text{Re} (\eta)$ will be given by
\begin{equation}\label{phase2}\gamma\simeq - \text{Im}\left({\lambda^{2}} \bigg[n\frac{C_{-,\alpha}}{k_{\alpha}L}+\! \sum_{\beta\neq \alpha} \frac{C_{+, \beta}^{*}}{k_{\beta}L} +  (n+1)\frac{C_{+,\alpha}^{*}}{k_{\alpha}L} \bigg]\right).\end{equation}

We now make two observations. Mode invisibility, which imposes that $\alpha$ is even, implies that $C_{-,\alpha}=0$ (See appendix \ref{cognac}). Additionally, since we want to compare the phase acquired by atoms crossing cavities with different photon content, we are interested in the difference between phases for different states containing $n$ and  $n+m$ which is given by  
{\begin{align}
\Delta_m\gamma(n)=&\gamma(n+m)-\gamma(n),
\end{align}}
which yields
 {\begin{equation}\label{phase3}
\Delta_m\gamma\simeq \frac{\lambda^{2}m}{k_{\alpha}L} \text{Im}\, C_{+,\alpha} 
\end{equation}}
 upon substituting \eqref{phase2}.  While  the complete expression calculated from \eqref{phasecal} does depend on $n$,  if we remain in a regime such that $\gamma\ll1$, then $\Delta_m \gamma$ is independent of $n$ within the approximation.

 Computing the value of the  integral $C_{+,\alpha}$ for non-relativistic speeds $v/c\ll1$ we get from  \eqref{Cfinal} 
\[\text{Im} \,C_{+,\alpha}\simeq \frac{-c }{\pi\alpha v}\frac{L^2}{v^2-4c^2}\simeq \frac{L^2 }{4\pi\alpha c v},\]
which upon substituting in \eqref{phase3} and taking into account that $k_\alpha L=\pi\alpha$ yields 
\begin{equation}\label{phase4}\Delta_m\gamma\simeq \frac{ \lambda^2 L^2}{4\pi^2\alpha^2 cv }m.\end{equation}
showing that in the very-few-photon regime the phase difference between two Fock states would not be a function of $n$ and would depend only on $m$. Finally, notice that the the exact expression for the phase difference is then
 \begin{align*}
\Delta_{m} \gamma(n) \!=\! \text{Re}\!\left[\ii \operatorname{Ln} \left(\frac{1 - \! \lambda^{2} \bigg[\! \sum \limits_{\beta\neq \alpha} \frac{C_{+, \beta}^{*}}{k_{\beta}L} +  \frac{(n+m+1)C_{+,\alpha}^{*}}{k_{\alpha}L} \bigg]}{1 - \! \lambda^{2} \bigg[\! \sum\limits_{\beta\neq \alpha} \frac{C_{+, \beta}^{*}}{k_{\beta}L} +  \frac{(n+1)C_{+,\alpha}^{*}}{k_{\alpha}L}\bigg]}\right)\right]
\end{align*}

\begin{figure}
\begin{center}
\includegraphics[width=.4\textwidth]{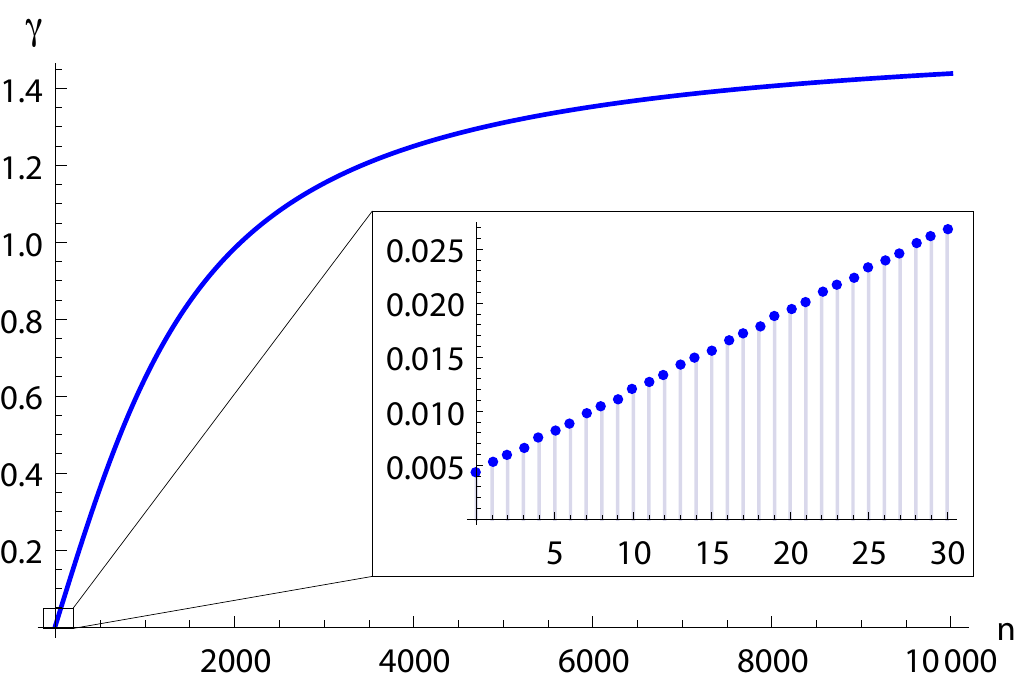}
\caption{(Color online): Plot of the phase acquired by a detector after it crosses the cavity (with speed $v=1000$ m/s) versus the number $n$ of photons  detected in the cavity (Real part of $\gamma$).  We are better able to distinguish between smaller numbers of photons than higher numbers of photons in the cavity. We see that as the number of photons increases, the slope of the curve decreases logarithmically,  worsening the resolution.}
\label{phase}
\end{center}
\end{figure}

As one can easily see, the relevant contribution to this phase comes from the resonant modes and the integrals $C_{+,\alpha}$. Unlike the leading order contribution to the probability of transition, the resonant contribution to the phase is not cancelled out by the `mode invisibility' technique. 

 Note that our calculations are made in the context of perturbation theory.  An estimator of the validity of the approximation 
 is  $\Omega^{-1}\lambda\cdot n\cdot(L/v) \ll 1$; when this criterion is satisfied, transition probabilities are small. For microwave cavities this give $10^{-9} n \ll 1$; for optical cavities the bound is even tighter.  However, when the phase obtained in our plots approaches order 1 (for the parameters considered that would happen around 1000 photons), one might question if  we are still within the perturbative regime. It can be checked that higher order corrections to $\gamma$ tend to increase the phase rather than decrease it; hence we expect our results extend to states of more photons without major problems. The visibility factor, however, would decrease substantially outside of the perturbative regime so there will always be a tradeoff imposed by the interferometric sensibility for cases where we want to probe a high enough number of photons.

\section{Results}

 To consider a particular case, we present results for cavities ranging from the microwave to the optical regime. From  \eqref{phasecal}  (and \eqref{Cfinal}) we see that if the coupling strength is chosen to be proportional to $\Omega$, which in turn is fixed to be on resonance with the $\alpha$-th harmonic of the cavity, the phase will be invariant under changes of length of the cavity provided we keep the same ratio $\lambda/\Omega$. If this ratio changes, the effect will be changed as the square of this magnitude. The coupling strength $\lambda$ for the microwave to the optical regime lies in the range $(10^{-6 } - 10^{-4})\Omega$ as is typical in quantum optical settings \cite{cavities}.

To consider a particular case, we present results for an optical microcavity of  length $L\sim1$ $\mu$m (although we know the results would be similar for a microwave cavity as per the arguments above).  We will consider the atomic gap to be resonant with a lower even harmonic of the cavity (whose spatial distribution is  odd).  In the relevant cavity mode, there is an unknown Fock state  whose photon content we want to determine. 

We consider additionally another cavity prepared with a reference field state and set up an atomic interferometer as shown in Fig.  \ref{atominterferometer}.  We need first to make sure that, as claimed,  the approximation \eqref{phased} holds when we send the atom with a given constant speed through the cavity due to the mode invisibility effect. We find that the   transition probability -- even for a relatively strong coupling $\lambda = 10^{-4}\Omega$ --  remains below $10^{-20}$ for our choice of parameters, consistent with our perturbative approach and the assumption \eqref{phased}.\footnote{This is similar to the rate of  response  in the case of vacuum fluctuations, since the mode that is populated has been made invisible to the detector,  as per the technique spelled out in section \ref{stealth}. }  Hence for realistic values of the  parameters the `mode invisibility' technique is rather effective: the atom will not significantly modify the state of the field while flying through it.

One may then wonder how much information can be gained about the state of the field. We show  in Fig. \ref{phase} the relationship between the phases acquired by a detector flying though the cavity and the number of photons contained in the relevant field mode. For very few photons, the phase response is linear with the number of photons. The curve deviates from linearity for large numbers of photons, reducing the resolution of the setting. This implies that our setting is better suited to distinguish between states  whose photon numbers differ by large amounts;  it is more difficult to distinguish a state containing $n+1$ photons from a state containing $n$ photons.

\begin{figure}
\begin{center}
\includegraphics[width=.45\textwidth]{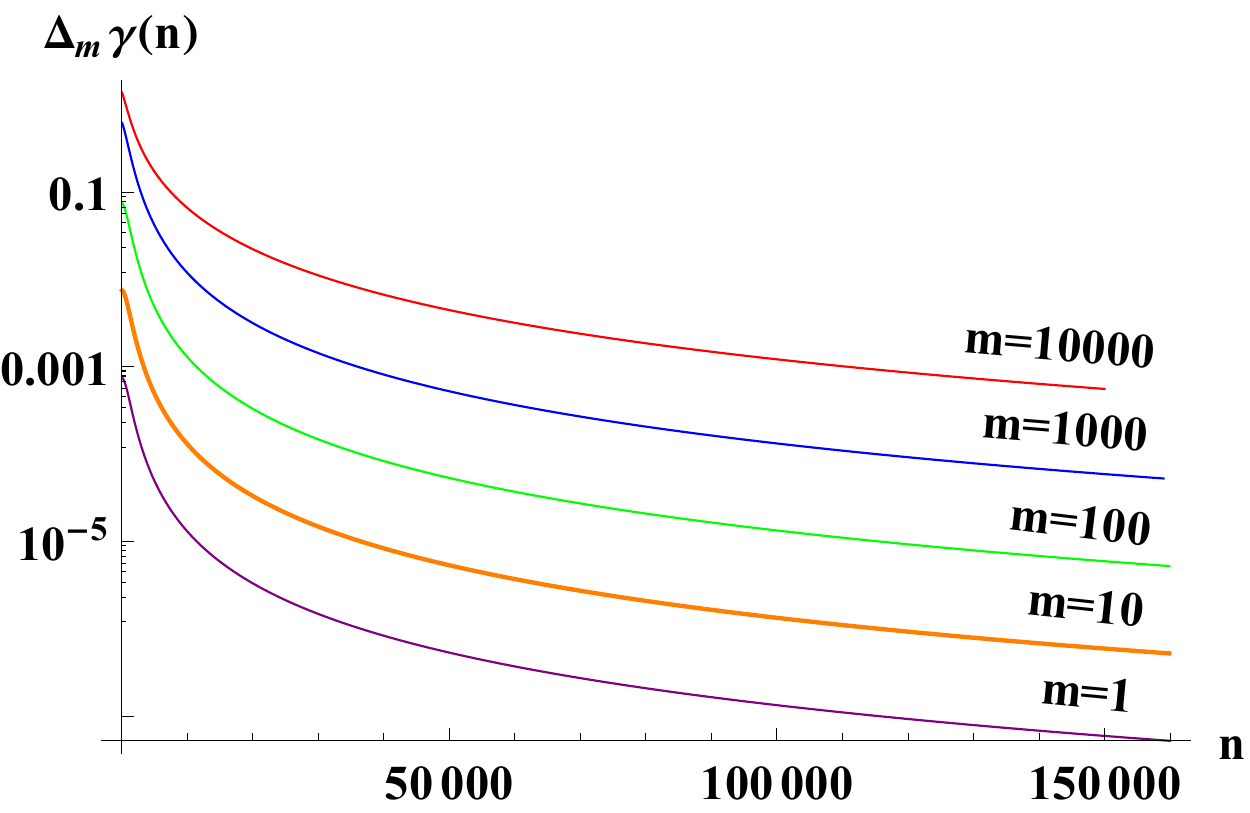}
\caption{(Color online): Phase resolution required to distinguish a state containing $n+m$ photons from a state containing  $n$ photons.  To show the trend in the behaviour of this magnitude it is interesting to plot it for $n$ above the threshold where the visibility would make the measurement experimentally challenging ($\sim10^4$ photons: see Fig \ref{visibilityfactor}) }
\label{phaseresolution}
\end{center}
\end{figure}

\begin{figure}
\begin{center}
\includegraphics[width=.45\textwidth]{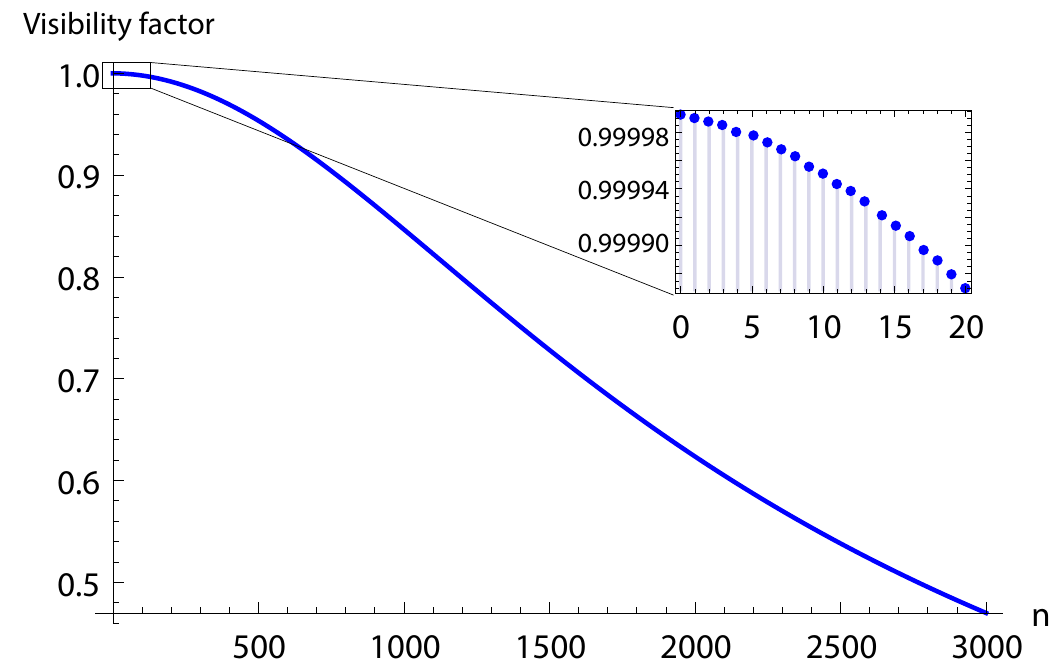}
\caption{(Color online): Plot of the visibility factor $e^{|\text{Im}{\eta}|}$  showing that the interferometry will not be significantly disturbed by the second order effects that take us out of the approximation \eqref{hypo}. Here $v=1000$ m/s as in the previous plots.  }
\label{visibilityfactor}
\end{center}
\end{figure}

The magnitude of the resolution of the interferometric experiment  conditions our ability to distinguish the number of photons in the unknown state of the field. Knowing that typical resolutions   in atomic interferometry are of the order of   fractions of milliradians \cite{cavities}, we see that for small numbers of photons we have more than enough resolution to tell apart states that differ by only one photon. The resolution  rapidly decreases with the number of photons.

Although we might not be able to tell apart a state with a million photons from a state of a million and one photons, we can still obtain information about the photon content of the unknown state.  The difference in phase between two states containing respectively $n$ and $n+m$ photons  is thus the required resolution of the interferometric experiment in order to distinguish states that differ by $m$ photons. We show $\Delta_m\gamma(n)$ in Fig. \ref{phaseresolution} for several values of $m$. 

The last point we need to analyze is how good the visibility of the interferometric pattern will be, taking into account that we would have second order effects taking us out of the ground state. They are guaranteed to be small due to their second order nature, but they would impact the visibility of the fringes. In Fig. \ref{visibilityfactor} we plot the visibility factor showing that the visibility remains extremely close to unity for small numbers of photons.

\section{Conclusions}
We have proposed a way to obtain the relative difference in the number of photons of two different Fock states without perturbing the states of the field. For this, we have made use of what we call the ``mode invisibility'' technique. This allows us to cancel the largest contribution to the transition probability  of an atom going through a photon-populated cavity while keeping the phase acquired by the atom sensibly high. In this way it is possible to obtain information about the quantum state of light without perturbing the system very much via an atomic interferometric experiment. For realistic values of the parameters (microwave cavities) there is enough resolution to distinguish states that differ only by one photon and whose photon population can be of the order of 1000.

 This opens up the possibility of constructing extremely sensitive measurement schemes with the ability to detect and identify states of light containing only a few photons with small measurement error.  The setting was presented here only for Fock states, but it could be applied to build more general settings where different families of states can be distinguished, as for instance coherent states, squeezed states, thermal states (photon content), fluctuations, etc.

\section{Acknowledgments}
This work was supported in part by the Natural Sciences and Engineering Research Council of Canada. E. M-M acknowledges support of the Banting Postdoctoral Fellowship Programme.  We are grateful to Keith Copsey for helpful comments.

\appendix
\begin{widetext}
\section{Evolution of the original quantum system}
Explicitly writing the interaction Hamiltonian in equation \eqref{interactionhamiltonian}:
\begin{align*}
H_{I} =&  \sum_{\gamma}\frac{\lambda}{\sqrt{k_{\gamma}L}}\bigg( \sigma^{+}a_{\gamma}^{\dagger}e^{\ii(\Omega + \omega_{\gamma})t} +  \sigma^{+}a_{\gamma}e^{-\ii(-\Omega + \omega_{\gamma})t}+ \sigma^{-} a_{\gamma}^{\dagger}e^{\ii(-\Omega + \omega_{\gamma})t}
+ \sigma^{-} a_{\gamma}e^{-\ii(\Omega + \omega_{\gamma})t} \bigg) \sin(k_{\gamma}x(t)).
\end{align*}
The state of our quantum system at time $t=0$ is defined as in equation \eqref{initialstate}. To find the system's state at later time $T$, we consider the evolution operator (see equation \eqref{unitary}) which we explicitly defined in equation \eqref{eq:pert}.
Before we proceed to calculations, it is important we define the action of annihilation and creation operators of a two-level system and bosonic field on the state of an atom and boson state respectively:
\begin{align*}
\hat{\sigma}^{-}|g\rangle &= 0, ~\hat{\sigma}^{+}|g\rangle = \ket{e}, ~\hat{\sigma}^{-}|e\rangle = \ket{g}, ~\hat{\sigma}^{+}|e\rangle = 0,\\
\hat{a}^{\dagger}_\beta\ket{n_\alpha} &= \sqrt{n+1}\ket{(n+1)_\alpha} \delta_{\beta\alpha},~\hat{a}_\beta \ket{n_\alpha} = \sqrt{n}\ket{(n-1)_\alpha}\delta_{\beta\alpha},~(\hat{a}_\beta^{\dagger})^{*} = \hat{a}_\beta.
\end{align*}
Therefore, given the initial quantum state $\ket{\psi(0)}=\ket{g}\otimes \ket{n_\alpha}\bigotimes_{\beta\neq \alpha}\ket{0_\beta}$, to first order in perturbation $\lambda$, 
\begin{align}\label{zero}
\nonumber\operatorname{U}^{(1)}\ket{\psi(0)}=& \int_{0}^{\frac{L}{v}} \text{dt}~ \sum_{\gamma}\frac{-\ii\lambda}{\sqrt{k_{\gamma}L}}\bigg( \sigma^{+}a_{\gamma}^{\dagger}e^{\ii(\Omega + \omega_{\gamma})t} +  \sigma^{+}a_{\gamma}e^{-\ii(-\Omega + \omega_{\gamma})t}\bigg) \ket{\psi(0)}\\
=& -\ii\frac{\lambda}{\sqrt{k_{\gamma}L}} \ket{e}\bigg[\sum_{\gamma \neq \alpha}X_{+,\gamma}\ket{n_{\alpha} 1_{\gamma}} + X_{+,\alpha}\sqrt{n + 1}\ket{(n+1)_{\alpha}}+  X_{-,\alpha}^{*}\sqrt{n}\ket{(n-1)_{\alpha}}\bigg],
\end{align}
Where for notational convenience, $X_{\pm, \gamma}$  has been defined as in equation \eqref{prob}. Equation \eqref{zero} gives a zero term contribution to the transition probability into a different state; That is $\operatorname{Tr}_{\text{F}}\rho(0) = 0$. Thus we go ahead to second order perturbation in $\lambda$.  Many terms emerge  for the unitary evolution operator $U^{(2)}$ to second order. However we concentrate only on surviving terms, which are
\begin{align*}
\operatorname{U}^{(2)}&=-\lambda^{2}\sum_{\gamma}\sum_{\beta}\int_{0}^{\frac{L}{v}}dt\int_{0}^{t}dt' {\Bigg( \sigma^{-}\sigma^{+} a^{\dagger}_{\gamma}a^{\dagger}_{\beta} e^{\ii(\omega_{\gamma} - \Omega)t} e^{\ii(\omega_{\beta} + \Omega)t'} + \sigma^{-}\sigma^{+} a^{\dagger}_{\gamma}a_{\beta} e^{\ii(\omega_{\gamma} - \Omega)t} e^{-\ii(\omega_{\beta} - \Omega)t'}} \\ 
& {+ \sigma^{-}\sigma^{+} a_{\gamma}a_{\beta} e^{-\ii(\omega_{\gamma} + \Omega)t} e^{-\ii(\omega_{\beta} - \Omega)t'}   +\sigma^{-}\sigma^{+} a_{\gamma}a^{\dagger}_{\beta} e^{-\ii(\omega_{\gamma} + \Omega)t} e^{\ii(\omega_{\beta} + \Omega)t'}\Bigg)} { \frac{\sin(k_{\gamma}x(t))}{\sqrt{k_{\gamma}L}}\frac{\sin(k_{\beta}{x(t'))}}{\sqrt{k_{\beta}L}}}.
\end{align*}
Therefore 
\begin{align*}
&U^{(2)}\ket{\psi(0)} =  \left(\frac{\lambda}{i}\right)^{2}\int_{0}^{\frac{L}{v}}dt\int_{0}^{t}dt'|g\rangle 
\Bigg[ \sum_{\gamma \neq \beta, \alpha}\sum_{\beta \neq \alpha} {\frac{\sin(k_{\gamma}x(t))}{\sqrt{k_{\gamma}L}} \frac{\sin(k_{\beta}x(t'))}{\sqrt{k_{\beta}L}} }e^{\ii(\omega_{\gamma} - \Omega)t} e^{\ii(\omega_{\beta} + \Omega)t'} |n_{\alpha}~1_{\beta}~1_{\gamma} \rangle \\
& + \frac{\sin(k_{\alpha}x(t)){\sin(k_{\alpha}x(t'))}}{k_{\alpha}L} e^{-\ii(\omega_{\alpha} + \Omega)(t-t')} (n+1)|n_{\alpha}\rangle +  \frac{\sin(k_{\alpha}x(t)) {\sin(k_{\alpha}x(t'))}}{k_{\alpha}L}e^{-\ii(\omega_{\alpha} + \Omega)(t+t')} \sqrt{n(n-1)}|(n-2)_{\alpha}\rangle
\\&+  \frac{\sin(k_{\alpha}x(t)){\sin(k_{\alpha}x(t'))}}{k_{\alpha}L} e^{\ii(\omega_{\alpha} - \Omega)(t-t')}  n|n_{\alpha}\rangle+ \sum_{\beta \neq \alpha} \frac{\sin(k_{\alpha}x(t))}{\sqrt{k_{\alpha}L}} \frac{{\sin(k_{\beta}x(t'))}}{\sqrt{k_{\beta}L}}e^{\ii(\omega_{\alpha} - \Omega)t} e^{\ii(\omega_{\beta} + \Omega)t'}\sqrt{n+1}|(n+1)_{\alpha}~1_{\beta}\rangle \\ 
 &{+ \sum_{\beta \neq \alpha} \frac{\sin(k_{\alpha}x(t))}{\sqrt{k_{\alpha}L}} \frac{\sin(k_{\beta}x(t'))}{\sqrt{k_{\beta}L}}e^{\ii(\omega_{\beta} - \Omega)t} e^{\ii(\omega_{\alpha} + \Omega)t'}\sqrt{n+1}|(n+1)_{\alpha}~1_{\beta}\rangle }\\ 
&{+ \sum_{\beta \neq \alpha} \frac{\sin(k_{\beta}x(t))}{\sqrt{k_{\beta}L}} \frac{\sin(k_{\alpha}x(t'))}{\sqrt{k_{\alpha}L}}e^{\ii(\omega_{\beta} - \Omega)t} e^{-\ii(\omega_{\alpha} - \Omega)t'} \sqrt{n}|(n-1)_{\alpha} 1_{\beta} \rangle} \\& + \sum_{\beta \neq \alpha}  \frac{\sin(k_{\beta}x(t)) \sin(k_{\beta}x(t'))}{k_{\beta}L} e^{-\ii(\omega_{\beta} + \Omega)(t-t')} |n_{\alpha}~0_{\beta}\rangle\\& + \sum_{\beta \neq \alpha} \frac{\sin(k_{\alpha}x(t)) \sin(k_{\alpha}x(t'))}{k_{\alpha}L}e^{\ii(\omega_{\alpha} - \Omega)t} e^{\ii(\omega_{\alpha} + \Omega)t'} \sqrt{(n+1)(n+2)}|(n+2)_{\alpha}\rangle\\
&  {+ \sum_{\beta \neq \alpha} \frac{\sin(k_{\alpha}x(t))}{\sqrt{k_{\alpha}L}} \frac{\sin(k_{\beta}x(t'))}{\sqrt{k_{\beta}L}}e^{-\ii(\omega_{\beta} - \Omega)t}e^{-\ii(\omega_{\alpha} - \Omega)t'}  \sqrt{n}|(n-1)_{\alpha}~1_{\beta}\rangle}\\ &+ \sum_{ \beta \neq \alpha}  \frac{\sin(k_{\beta}x(t)) \sin(k_{\beta} x(t'))}{k_{\beta}L}e^{\ii(\omega_{\beta} - \Omega)t} e^{i(\omega_{\beta} + \Omega)t'}\!\!\sqrt{2}|n_{\alpha}~2_{\beta}\rangle  \Bigg].
\end{align*}
We therefore have to second order  in $\lambda$  the state
\begin{align}\label{secondorder}
\ket{\psi^{(2)}(T)} = &\frac{-\lambda^{2}}{k_{\alpha}L} \bigg( C_{+,\alpha}^{*}(n+1)+ C_{-,\alpha} n + \sum_{\beta \neq \alpha}  C_{+,\alpha}^{*} \bigg)\ket{\psi(0)}+ \ket{\psi(T)}_{\perp},
\end{align}
where the last term represents other terms orthogonal to the initial quantum state and $C_{\pm,\beta}$ has been defined as in equation \eqref{crossterms}.

\section{Transition Probability}\label{anything}

Given that our leading order in the perturbation theory is the second order, the transition probability of an atom initially in its ground state into an excited state is given as {$\bra{e}\operatorname{Tr}_f\big[U^{(1)}\rho_{0}U^{(1)\dagger}+U^{(2)} \rho_{0} + U^{(2)\dagger}\rho_0\big]\ket{e}$} where $\rho_{0} = \ket{\psi(0)}\langle \psi(0)|$. {Notice that the summands including $U^{(2)}$ do not produce diagonal terms in the field, hence they will not contribute to the transition probability.}
\begin{align*}
\bigg(U^{(1)}\ket{\psi(0)}\bigg)\bigg(\langle \psi(0)|U^{(1)\dagger}\bigg) = & \left(\frac{\lambda}{\ii}\right)|e\rangle \langle g| \bigg[\sum_{\gamma \neq \alpha}X_{+,\gamma}|n_{\alpha} 1_{\gamma}\rangle\langle n_{\alpha }0_{\gamma} | + X_{+,\alpha}\sqrt{n + 1}|(n+1)_{\alpha}\rangle\langle n_{\alpha} | \\
& \qquad \qquad \qquad +  X_{-,\alpha}^{*}\sqrt{n}|(n-1)_{\alpha} \rangle \langle n_{\alpha}|  \left(\frac{-\lambda}{\ii}\right)\sum_{\beta}\bigg( \sigma^{-}a_{\beta}X_{+,\beta}^{*} + \sigma^{-}a_{\beta}^{\dagger}X_{-,\beta}\bigg)\bigg].
\end{align*}
Taking each term explicitly, 
\begin{align*}
&\lambda^{2}|e\rangle \langle g| \bigg[\sum_{\gamma \neq \alpha}X_{+,\gamma}|n_{\alpha} 1_{\gamma}\rangle\langle n_{\alpha }~0_{\gamma} | + X_{+,\alpha}\sqrt{n + 1}|(n+1)_{\alpha}\rangle\langle n_{\alpha} | +  X_{-,\alpha}^{*}\sqrt{n}|(n-1)_{\alpha} \rangle \langle n_{\alpha}| \bigg]\sum_{\beta}X_{-,\beta}\sigma^{-}a_{\beta}^{\dagger}\\
=& \lambda^{2} |e\rangle\langle e|\Bigg[\sum_{\gamma \neq     \alpha}X_{+,\gamma}X_{-\alpha} \sqrt{n}|n_{\alpha}~1_{\gamma}\rangle \langle (n-1)_{\alpha}~ 0_{\gamma}|+X_{+,\alpha}X_{-~\alpha} \sqrt{n(n+1)}|(n+1)_{\alpha}\rangle \langle (n-1)_{\alpha}| \\
&\qquad\qquad\qquad+ |X_{-\alpha}|^{2}n|(n-1)_\alpha \rangle \langle (n -1)_{\alpha}|\Bigg].
\end{align*}
Similarly for terms in $\sigma^{-}a_{\beta}X_{+,\beta}^{*}$, we have
\begin{align*}
&\left(\frac{\lambda}{i}\right)^{2}|e\rangle \langle g| \bigg[\sum_{\gamma \neq \alpha}X_{+,\gamma}|n_{\alpha} 1_{\gamma}\rangle\langle n_{\alpha }~0_{\gamma} | + X_{+,\alpha}\sqrt{n + 1}|(n+1)_{\alpha}\rangle\langle n_{\alpha} | +X_{-,\alpha}^{*}\sqrt{n}|(n-1)_{\alpha} \rangle \langle n_{\alpha}| \bigg]\sum_{\beta}X_{+,\beta}^{*}\sigma^{-}a_{\beta}\\
=& \left(\frac{\lambda}{i}\right)^{2}|e\rangle \langle e| \Bigg[\sum_{\beta \neq \gamma,\alpha}\sum_{ \gamma \neq \alpha}X_{+,\gamma}X_{+,\beta}^{*}|n_{\alpha}~ 1_{\gamma}~0_{\beta}\rangle \langle n_{\alpha}~0_{\gamma}~1_{\beta}| + \sum_{\gamma \neq\alpha }X_{+,\gamma}X_{+,\alpha}^{*}\sqrt{n+1}|n_{\alpha}~ 1_{\gamma}\rangle \langle (n+1)_{\alpha} ~0_{\gamma} | \\
&   + \sum_{\gamma \neq \alpha}|X_{+,\gamma}|^{2} |n_{\alpha}~1_{\gamma} \rangle \langle n_{\alpha} 1_{\gamma} | + \sum_{\beta \neq \alpha} X_{+,\alpha}X_{+,\beta}^{*} \sqrt{n+1}|(n+1)_{\alpha}~0_{\beta}\rangle \langle n_{\alpha}~1_{\beta} | + |X_{+,\alpha}|^{2} (n+1)|(n+1)_{\alpha}\rangle \langle (n+1)_{\alpha}|\\
&   + \sum_{\beta\neq \alpha} X_{-, \alpha}^{*}X_{+, \beta}^{*} \sqrt{n}|(n-1)_{\alpha}~0_{\beta}\rangle \langle n_{\alpha}~1_{\beta}|
+X_{-,\alpha}^{*}X_{+,\alpha}^{*} \sqrt{n(n+1)}|(n-1)_{\alpha}\rangle \langle (n+1)_{\alpha}| \Bigg].
\end{align*}
Combining these terms, we have
\begin{align*}
U^{(1)} \rho_{0} U^{(1)^{\dagger}} = &\lambda^{2}|e\rangle \langle e|\Bigg[\sum_{\gamma \neq \alpha}X_{+\gamma}X_{-\alpha} \sqrt{n}|n_{\alpha}~1_{\gamma}\rangle \langle (n-1)_{\alpha}~ 0_{\gamma}|+X_{+,\alpha}X_{-~\alpha} \sqrt{n(n+1)}|(n+1)_{\alpha}\rangle \langle (n-1)_{\alpha}| \\
&\qquad  \qquad + |X_{-\alpha}|^{2}n|n-1_\alpha \rangle \langle n -1_{\alpha}| + \sum_{\beta \neq \gamma,\alpha}\sum_{\gamma \neq \alpha}X_{+,\gamma}X_{+,\beta}^{*}|n_{\alpha}~ 1_{\gamma}~0_{\beta}\rangle \langle n_{\alpha}~0_{\gamma}~1_{\beta}|\\
& \qquad  \qquad +\sum_{ \gamma \neq \alpha}X_{+,\gamma}X_{+,\alpha}^{*}\sqrt{n+1}|n_{\alpha}~ 1_{\gamma}\rangle \langle (n+1)_{\alpha} ~0_{\gamma} | + \sum_{\gamma \neq \alpha}|X_{+,\gamma}|^{2} |n_{\alpha}~1_{\gamma} \rangle \langle n_{\alpha} ~1_{\gamma} |\\
& \qquad \qquad +\sum_{\beta \neq \alpha}X_{+,\alpha}X_{+,\beta}^{*} \sqrt{n+1}|(n+1)_{\alpha}~0_{\beta}\rangle \langle n_{\alpha}~1_{\beta} |
+ |X_{+,\alpha}|^{2} (n+1)|(n+1)_{\alpha}\rangle \langle (n+1)_{\alpha}|\\
&\qquad  \qquad +\sum_{\beta\neq \alpha}X_{-, \alpha}^{*}X_{+, \beta}^{*} \sqrt{n}|(n-1)_{\alpha}~0_{\beta}\rangle \langle n_{\alpha}~1_{\beta}|
+X_{-,\alpha}^{*}X_{+,\alpha}^{*} \sqrt{n(n+1)}|(n-1)_{\alpha}\rangle \langle (n+1)_{\alpha}| \Bigg].
\end{align*}
Tracing over the field, we have the transition probability of an atom into an excited state as seen in equation \eqref{transitionp}
\begin{align}
P_{\ket{e}}(T) = {\bra{e}}\operatorname{Tr}_{f}\bigg[U^{(1)} \rho_{0} U^{(1)^{\dagger}} \bigg] {{\ket{e}}} = \lambda^{2}\Bigg(\big|X_{-,\alpha}\big|^{2}n ~+~ \sum_{\gamma\neq \alpha} \big|X_{+, \gamma}\big|^{2} ~+ ~\big|X_{+,\alpha}\big|^{2}(n+1)\Bigg).
\end{align}
The integral
\begin{align*}
X_{\pm,\beta} = &\frac{1}{\sqrt{k_{\beta}L}}\int_{0}^{L/v}dt e^{\ii(\pm\Omega + \omega_{\beta})t}\sin[k_{\beta}vt] \\
\end{align*}
can be easily evaluated integrating by parts, yielding expression \eqref{inv2}.  Its modulus squared is evaluated to yield
\begin{align}\label{gerome}
\nonumber |X_{\pm,\beta}|^{2} =&\frac{1}{(k_{\beta} L)\Big[\Big(\frac{\pi \beta {v}}{L}\Big)^{2} -\Big(\pm\Omega + \frac{\pi \beta c}{L}\Big)^2\Big]^{2}}\Bigg[ \Big(\pm \Omega +\frac{\pi c \beta}{L}\Big)^{2}\sin^{2}(\pi \beta) +  \Big(\frac{\pi \beta}{L}v\Big)^{2}\cos^{2}(\pi \beta)  + \Big(\frac{\pi \beta}{L}v\Big)^2 \\
 &- 2\Big(\frac{\pi \beta v}{L}\Big)\Big(\pm\Omega + \frac{\pi \beta c}{L}\Big)\sin(\pi \beta)\sin\Big(\frac{\pm\Omega L}{v} + \frac{\pi \beta c}{v}\Big)- 2\Big( \frac{\pi \beta v}{L}\Big)^{2}\cos(\pi \beta)\cos\Big(\frac{\pm\Omega L}{v} + \frac{\pi \beta c}{v}\Big)\Bigg].
\end{align}
It is trivial to show that at resonance ($\omega_{\alpha} = \Omega), ~ |X_{-,\alpha}|^{2} = 0 $ while $|X_{+,\alpha}|^{2} $ is given by
\begin{align*}
|X_{+,\alpha}|^{2} 
= \frac{\Big( \frac{\pi \alpha v}{L}\Big)^{2} \sin^2\Big(\frac{\pi \alpha c}{v}\Big) }{(\pi \alpha)\Big[\Big(\frac{\pi \alpha {v}}{L}\Big)^{2} -\Big( \frac{2\pi \alpha c}{L}\Big)^2\Big]^{2}}.
\end{align*}
for $\alpha = 2j$ an even integer.  Provided 
\begin{align*}
\sin\Big(\frac{\pi \alpha c}{v}\Big) = 1 \Longrightarrow \quad v = \frac{2 j c}{N},
\end{align*}
where $N$ is an integer, we can be sure that the cavity content is completely invisible to the atom during its motion through the cavity.

 \section{Evaluating the phase factor}\label{cognac}
 
We have established an expression for the state of the quantum system after the time evolution operation to be $| \psi (T)\rangle = \ket{\psi(0)} +  \ket{\psi}_{T}^{(2)}+ \cdots$. Substituting equation \eqref{secondorder} yields,
\begin{align}\label{third}
| \psi (T)\rangle =& \ket{\psi(0)} + \frac{\lambda^{2}}{k_{\alpha}L} \Big[ n C_{-,\alpha} + \sum_{\beta \neq \alpha}C_{+,\beta}^{*}  + (n+1)C_{+,\alpha}^{*}\Big]\ket{\psi(T)} +\ket{\psi(T)}_\perp.
\end{align}
From equation \eqref{phasecal}, we can compute the integral term of the form
\begin{align*}
C_{\pm, \beta} = &\int\limits_{0}^{\frac{L}{v}} \mathrm{d}t \int\limits_{0}^{t}\mathrm{dt'}~e^{\ii(\omega_{\beta} \pm \Omega)t} ~ e^{-\ii(\omega_{\beta} \pm \Omega)t'}\sin[k_{\beta}x(t)]\sin[k_{\beta}x(t')]\\
= -& \int\limits_{0}^{\frac{L}{v}} \mathrm{d}t \frac{e^{\ii(\omega_{\beta} \pm \Omega)t}\sin[k_{\beta}x(t)]}{\Big[(k_{\beta}v)^{2} -(\pm \Omega + \omega_{\beta})^2\Big] } \Bigg[\ii(\pm \Omega + \omega_{\beta})\sin(k_{\beta}vt') +({k_{\beta}}v)\cos(k_{\beta}vt')\Bigg] e^{-\ii(\pm \Omega + \omega_{\beta})t'}\Bigg|_{0}^{t} \\ 
= &  - \int\limits_{0}^{\frac{L}{v}}  \frac{e^{\ii(\omega_{\beta} \pm \Omega)t}\sin[k_{\beta}x(t)]}{\Big[(k_{\beta}v)^{2} -(\pm \Omega + \omega_{\beta})^2\Big] } \Bigg[\ii(\pm \Omega + \omega_{\beta})\sin(k_{\beta}vt)e^{-\ii(\pm \Omega + \omega_{\beta})t } - (k_{\beta}v) +
(k_{\beta}v)\cos(k_{\beta}vt) e^{-\ii(\pm \Omega + \omega_{\beta})t } \Bigg]
\end{align*}
{Separating every summand we get}
\begin{align}
C_{\pm, \beta}=&\underbrace{\frac{ -\ii(\pm \Omega + \omega_{\beta})}{\Big[(k_{\beta}v)^{2} -(\pm \Omega + \omega_{\beta})^2\Big] } \int\limits_{0}^{\frac{L}{v}} \mathrm{dt}~ \sin[k_{\beta}vt]\sin(k_{\beta}vt) }_{P}\\
&  - \underbrace{\frac{(k_{\beta}v)}{\Big[(k_{\beta}v)^{2} -(\pm \Omega + \omega_{\beta})^2\Big] } \int\limits_{0}^{\frac{L}{v}} \mathrm{dt}~ \sin[k_{\beta}x(t)]\cos(k_{\beta}vt)}_{Q} \\
&+\underbrace{ \frac{(k_{\beta}v)}{\Big[(k_{\beta}v)^{2} -(\pm \Omega + \omega_{\beta})^2\Big] } \int\limits_{0}^{\frac{L}{v}} \mathrm{dt}~e^{\ii(\omega_{\beta} \pm \Omega)t} \sin[k_{\beta}x(t)]}_{R}.
\end{align}
Applying the method of integration as above yields
\begin{align*}
P =&  -   \frac{\ii L(\pm\Omega + \omega_{\beta})}{2v\Big[(k_{\beta}v)^{2} -(\Omega + \omega_{\beta})^2\Big] },\\
Q=&  \frac{\cos(2k_{\beta}L) }{4\Big[(k_{\beta}v)^{2} -(\pm\Omega + \omega_{\beta})^2\Big]} -\frac{1}{4\Big[(k_{\beta}v)^{2} -(\pm\Omega + \omega_{\beta})^2\Big]}, \\
R =&     \frac{(k_{\beta} v)^{2}\Big[\cos(k_{\beta}L)e^{\ii (\Omega + \omega_{\beta})L/v}-1 \bigg]}{\Big[(k_{\beta}v)^{2} -(\Omega + \omega_{\beta})^2\Big]^{2}}.
\end{align*}
On substituting $k_\beta = \pi\beta/L$ and $\omega_\beta = \pi\beta c/L$ we obtain 
\begin{align*}
C_{\pm,\beta} = & - \frac{\ii L(\pm\Omega +\pi\beta c/L)}{2v\Big[(\pi\beta v/L)^{2} -(\pm \Omega + \pi\beta c/L)^2\Big] }+\frac{(\pi\beta v/L)^{2}\Big[(-1)^{\beta}e^{\ii(\pm\Omega + \pi\beta c/L)L/v}  - 1\Big]}{\Big[(\pi\beta v/L)^{2} -(\pm\Omega + \pi\beta c/L)^2\Big]^{2}}.
\end{align*}

If  $\Omega$ is such as we are in resonance with the $\alpha$-th mode of the cavity, this is $\Omega=\frac{\alpha \pi}{L}c$,  $C_{\pm,\beta}$ can be expressed as
\begin{align}\label{Cfinal}
C_{\pm, \beta}=L^2\left(\frac{ (\beta \pi v)^{2}\left[(-1)^{\beta}e^{\ii[\pi\frac{ c}{v} (\beta \pm\alpha )]}-1\right]  }{\Big[(\beta \pi v)^{2} -\pi^2c^2(\beta \pm \alpha )^2\Big]^{2} }-  \frac{\ii \pi c (\beta \pm \alpha )}{2 v\Big[(\beta \pi v)^{2} -\pi^2c^2(\beta \pm \alpha )^2\Big] }\right).  
\end{align}
In particular, if $\beta=\alpha$ and if $\alpha$ is even, then $C_{-,\alpha}=0$ and
\begin{align}\label{Cfinal2}
C_{+, \alpha}=\frac{L^2}{\pi^2\alpha^2 v^2}\frac{ e^{i(2\pi\alpha\frac{ c}{v} )}-1  }{\Big(1  -4\frac{c^2}{v^2}\Big)^{2}  }-\frac{\ii c }{\pi\alpha v}\frac{L^2}{v^2-4c^2},
\end{align}
which in the non-relativistic limit $v/c\ll1$ transforms into
\begin{align}\label{Cfinal3}
{C_{+, \alpha}\simeq\frac{\ii\, L^2 }{4\pi\alpha c v}}
\end{align}
and so  the phase \eqref{phasecal} which only depends on the $C_{\pm,\alpha}$ is $L$-invariant, indicating that our method is scalable for any size of the cavity.

\end{widetext}

\bibliography{references}

\begin{thebibliography}{17}%
\makeatletter
\providecommand \@ifxundefined [1]{%
 \@ifx{#1\undefined}
}%
\providecommand \@ifnum [1]{%
 \ifnum #1\expandafter \@firstoftwo
 \else \expandafter \@secondoftwo
 \fi
}%
\providecommand \@ifx [1]{%
 \ifx #1\expandafter \@firstoftwo
 \else \expandafter \@secondoftwo
 \fi
}%
\providecommand \natexlab [1]{#1}%
\providecommand \enquote  [1]{``#1''}%
\providecommand \bibnamefont  [1]{#1}%
\providecommand \bibfnamefont [1]{#1}%
\providecommand \citenamefont [1]{#1}%
\providecommand \href@noop [0]{\@secondoftwo}%
\providecommand \href [0]{\begingroup \@sanitize@url \@href}%
\providecommand \@href[1]{\@@startlink{#1}\@@href}%
\providecommand \@@href[1]{\endgroup#1\@@endlink}%
\providecommand \@sanitize@url [0]{\catcode `\\12\catcode `\$12\catcode
  `\&12\catcode `\#12\catcode `\^12\catcode `\_12\catcode `\%12\relax}%
\providecommand \@@startlink[1]{}%
\providecommand \@@endlink[0]{}%
\providecommand \url  [0]{\begingroup\@sanitize@url \@url }%
\providecommand \@url [1]{\endgroup\@href {#1}{\urlprefix }}%
\providecommand \urlprefix  [0]{URL }%
\providecommand \Eprint [0]{\href }%
\providecommand \doibase [0]{http://dx.doi.org/}%
\providecommand \selectlanguage [0]{\@gobble}%
\providecommand \bibinfo  [0]{\@secondoftwo}%
\providecommand \bibfield  [0]{\@secondoftwo}%
\providecommand \translation [1]{[#1]}%
\providecommand \BibitemOpen [0]{}%
\providecommand \bibitemStop [0]{}%
\providecommand \bibitemNoStop [0]{.\EOS\space}%
\providecommand \EOS [0]{\spacefactor3000\relax}%
\providecommand \BibitemShut  [1]{\csname bibitem#1\endcsname}%
\let\auto@bib@innerbib\@empty
\bibitem [{\citenamefont {Braginsky}\ \emph {et~al.}(1977)\citenamefont
  {Braginsky}, \citenamefont {Vorontsov},\ and\ \citenamefont
  {Khalili}}]{Braginski}%
  \BibitemOpen
  \bibfield  {author} {\bibinfo {author} {\bibfnamefont {V.~B.}\ \bibnamefont
  {Braginsky}}, \bibinfo {author} {\bibfnamefont {Y.~I.}\ \bibnamefont
  {Vorontsov}}, \ and\ \bibinfo {author} {\bibfnamefont {F.~Y.}\ \bibnamefont
  {Khalili}},\ }\href@noop {} {\bibfield  {journal} {\bibinfo  {journal} {Zh.
  Eksp. Teor. Fiz.}\ }\textbf {\bibinfo {volume} {73}},\ \bibinfo {pages}
  {1340} (\bibinfo {year} {1977})}\BibitemShut {NoStop}%
\bibitem [{\citenamefont {Roch}\ \emph {et~al.}(1992)\citenamefont {Roch},
  \citenamefont {Roger}, \citenamefont {Grangier}, \citenamefont {Courty},\
  and\ \citenamefont {Reynaud}}]{firstdemo}%
  \BibitemOpen
  \bibfield  {author} {\bibinfo {author} {\bibfnamefont {J.~F.}\ \bibnamefont
  {Roch}}, \bibinfo {author} {\bibfnamefont {G.}~\bibnamefont {Roger}},
  \bibinfo {author} {\bibfnamefont {P.}~\bibnamefont {Grangier}}, \bibinfo
  {author} {\bibfnamefont {J.~M.}\ \bibnamefont {Courty}}, \ and\ \bibinfo
  {author} {\bibfnamefont {S.}~\bibnamefont {Reynaud}},\ }\href@noop {}
  {\bibfield  {journal} {\bibinfo  {journal} {Appl. Phys, B.}\ }\textbf
  {\bibinfo {volume} {55}},\ \bibinfo {pages} {291} (\bibinfo {year}
  {1992})}\BibitemShut {NoStop}%
\bibitem [{\citenamefont {Grangier}\ \emph {et~al.}(1998)\citenamefont
  {Grangier}, \citenamefont {Levenson},\ and\ \citenamefont
  {Poizat}}]{secondemo}%
  \BibitemOpen
  \bibfield  {author} {\bibinfo {author} {\bibfnamefont {P.}~\bibnamefont
  {Grangier}}, \bibinfo {author} {\bibfnamefont {J.~A.}\ \bibnamefont
  {Levenson}}, \ and\ \bibinfo {author} {\bibfnamefont {J.~P.}\ \bibnamefont
  {Poizat}},\ }\href@noop {} {\bibfield  {journal} {\bibinfo  {journal}
  {Nature}\ }\textbf {\bibinfo {volume} {396}},\ \bibinfo {pages} {537}
  (\bibinfo {year} {1998})}\BibitemShut {NoStop}%
\bibitem [{\citenamefont {Brune}\ \emph {et~al.}(1992)\citenamefont {Brune},
  \citenamefont {Haroche},\ and\ \citenamefont {Raimond}}]{Haroche}%
  \BibitemOpen
  \bibfield  {author} {\bibinfo {author} {\bibfnamefont {M.}~\bibnamefont
  {Brune}}, \bibinfo {author} {\bibfnamefont {S.}~\bibnamefont {Haroche}}, \
  and\ \bibinfo {author} {\bibfnamefont {J.~M.}\ \bibnamefont {Raimond}},\
  }\href@noop {} {\bibfield  {journal} {\bibinfo  {journal} {Am J Phys.}\
  }\textbf {\bibinfo {volume} {45}},\ \bibinfo {pages} {5193} (\bibinfo {year}
  {1992})}\BibitemShut {NoStop}%
\bibitem [{\citenamefont {Haroche}(2013)}]{Sergelecture}%
  \BibitemOpen
  \bibfield  {author} {\bibinfo {author} {\bibfnamefont {S.}~\bibnamefont
  {Haroche}},\ }\href@noop {} {\bibfield  {journal} {\bibinfo  {journal}
  {Reviews of Modern Physics}\ }\textbf {\bibinfo {volume} {85}},\ \bibinfo
  {pages} {1083} (\bibinfo {year} {2013})}\BibitemShut {NoStop}%
\bibitem [{\citenamefont {Rideout}\ \emph {et~al.}(2012)\citenamefont
  {Rideout}, \citenamefont {Jennewein}, \citenamefont {Amelino-Camelia},
  \citenamefont {Demarie}, \citenamefont {Higgins}, \citenamefont {Kempf},
  \citenamefont {Kent}, \citenamefont {Laflamme}, \citenamefont {Ma},
  \citenamefont {Mann}, \citenamefont {Mart{\'\i}n-Mart{\'\i}nez},
  \citenamefont {Menicucci}, \citenamefont {Moffat}, \citenamefont {Simon},
  \citenamefont {Sorkin}, \citenamefont {Smolin},\ and\ \citenamefont
  {Terno}}]{Rideout2012}%
  \BibitemOpen
  \bibfield  {author} {\bibinfo {author} {\bibfnamefont {D.}~\bibnamefont
  {Rideout}}, \bibinfo {author} {\bibfnamefont {T.}~\bibnamefont {Jennewein}},
  \bibinfo {author} {\bibfnamefont {G.}~\bibnamefont {Amelino-Camelia}},
  \bibinfo {author} {\bibfnamefont {T.~F.}\ \bibnamefont {Demarie}}, \bibinfo
  {author} {\bibfnamefont {B.~L.}\ \bibnamefont {Higgins}}, \bibinfo {author}
  {\bibfnamefont {A.}~\bibnamefont {Kempf}}, \bibinfo {author} {\bibfnamefont
  {A.}~\bibnamefont {Kent}}, \bibinfo {author} {\bibfnamefont {R.}~\bibnamefont
  {Laflamme}}, \bibinfo {author} {\bibfnamefont {X.}~\bibnamefont {Ma}},
  \bibinfo {author} {\bibfnamefont {R.~B.}\ \bibnamefont {Mann}}, \bibinfo
  {author} {\bibfnamefont {E.}~\bibnamefont {Mart{\'\i}n-Mart{\'\i}nez}},
  \bibinfo {author} {\bibfnamefont {N.~C.}\ \bibnamefont {Menicucci}}, \bibinfo
  {author} {\bibfnamefont {J.}~\bibnamefont {Moffat}}, \bibinfo {author}
  {\bibfnamefont {C.}~\bibnamefont {Simon}}, \bibinfo {author} {\bibfnamefont
  {R.}~\bibnamefont {Sorkin}}, \bibinfo {author} {\bibfnamefont
  {L.}~\bibnamefont {Smolin}}, \ and\ \bibinfo {author} {\bibfnamefont {D.~R.}\
  \bibnamefont {Terno}},\ }\href
  {http://stacks.iop.org/0264-9381/29/i=22/a=224011} {\bibfield  {journal}
  {\bibinfo  {journal} {Classical and Quantum Gravity}\ }\textbf {\bibinfo
  {volume} {29}},\ \bibinfo {pages} {224011} (\bibinfo {year}
  {2012})}\BibitemShut {NoStop}%
\bibitem [{\citenamefont {Mart\'{i}n-Mart\'{i}nez}\ \emph
  {et~al.}(2013{\natexlab{a}})\citenamefont {Mart\'{i}n-Mart\'{i}nez},
  \citenamefont {Fuentes},\ and\ \citenamefont {Mann}}]{Quantumthermometer}%
  \BibitemOpen
  \bibfield  {author} {\bibinfo {author} {\bibfnamefont {E.}~\bibnamefont
  {Mart\'{i}n-Mart\'{i}nez}}, \bibinfo {author} {\bibfnamefont
  {I.}~\bibnamefont {Fuentes}}, \ and\ \bibinfo {author} {\bibfnamefont
  {R.~B.}\ \bibnamefont {Mann}},\ }\href@noop {} {\bibfield  {journal}
  {\bibinfo  {journal} {New J. Phys.}\ }\textbf {\bibinfo {volume} {15}},\
  \bibinfo {pages} {1} (\bibinfo {year} {2013}{\natexlab{a}})}\BibitemShut
  {NoStop}%
\bibitem [{\citenamefont {Mart\'{i}n-Mart\'{i}nez}\ \emph
  {et~al.}(2011)\citenamefont {Mart\'{i}n-Mart\'{i}nez}, \citenamefont
  {Fuentes},\ and\ \citenamefont {Mann}}]{Unruh}%
  \BibitemOpen
  \bibfield  {author} {\bibinfo {author} {\bibfnamefont {E.}~\bibnamefont
  {Mart\'{i}n-Mart\'{i}nez}}, \bibinfo {author} {\bibfnamefont
  {I.}~\bibnamefont {Fuentes}}, \ and\ \bibinfo {author} {\bibfnamefont
  {R.~B.}\ \bibnamefont {Mann}},\ }\href@noop {} {\bibfield  {journal}
  {\bibinfo  {journal} {Phys. Rev. Lett.}\ }\textbf {\bibinfo {volume} {107}},\
  \bibinfo {pages} {1} (\bibinfo {year} {2011})}\BibitemShut {NoStop}%
\bibitem [{\citenamefont {Jonsson}\ \emph {et~al.}(2013)\citenamefont
  {Jonsson}, \citenamefont {Martin-Martinez},\ and\ \citenamefont
  {Kempf}}]{Robert2013}%
  \BibitemOpen
  \bibfield  {author} {\bibinfo {author} {\bibfnamefont {R.~H.}\ \bibnamefont
  {Jonsson}}, \bibinfo {author} {\bibfnamefont {E.}~\bibnamefont
  {Martin-Martinez}}, \ and\ \bibinfo {author} {\bibfnamefont {A.}~\bibnamefont
  {Kempf}},\ }\href@noop {} {\  (\bibinfo {year} {2013})},\ \bibinfo {note}
  {arXiv:1306.4275 [quant-ph]}\BibitemShut {NoStop}%
\bibitem [{\citenamefont {Scully}\ and\ \citenamefont
  {Zubairy}(1997)}]{ScullyBook}%
  \BibitemOpen
  \bibfield  {author} {\bibinfo {author} {\bibfnamefont {M.~O.}\ \bibnamefont
  {Scully}}\ and\ \bibinfo {author} {\bibfnamefont {M.~S.}\ \bibnamefont
  {Zubairy}},\ }\href@noop {} {\emph {\bibinfo {title} {Quantum Optics}}}\
  (\bibinfo  {publisher} {Cambridge University Press},\ \bibinfo {year}
  {1997})\BibitemShut {NoStop}%
\bibitem [{\citenamefont {Haroche}\ and\ \citenamefont
  {Raimond}(2006)}]{cavities}%
  \BibitemOpen
  \bibfield  {author} {\bibinfo {author} {\bibfnamefont {S.}~\bibnamefont
  {Haroche}}\ and\ \bibinfo {author} {\bibfnamefont {J.-M.}\ \bibnamefont
  {Raimond}},\ }\href@noop {} {\emph {\bibinfo {title} {Exploring the Quantum
  \textit{Atoms, Cavities, and Photons}}}}\ (\bibinfo  {publisher} {Oxford
  University Press},\ \bibinfo {year} {2006})\BibitemShut {NoStop}%
\bibitem [{\citenamefont {Raimond}\ \emph {et~al.}(2001)\citenamefont
  {Raimond}, \citenamefont {Brune},\ and\ \citenamefont {Haroche}}]{nature}%
  \BibitemOpen
  \bibfield  {author} {\bibinfo {author} {\bibfnamefont {J.~M.}\ \bibnamefont
  {Raimond}}, \bibinfo {author} {\bibfnamefont {M.}~\bibnamefont {Brune}}, \
  and\ \bibinfo {author} {\bibfnamefont {S.}~\bibnamefont {Haroche}},\
  }\href@noop {} {\bibfield  {journal} {\bibinfo  {journal} {Rev. Mod. Phys.}\
  }\textbf {\bibinfo {volume} {73}},\ \bibinfo {pages} {565} (\bibinfo {year}
  {2001})}\BibitemShut {NoStop}%
\bibitem [{\citenamefont {Mart\'{i}n-Mart\'{i}nez}\ \emph
  {et~al.}(2013{\natexlab{b}})\citenamefont {Mart\'{i}n-Mart\'{i}nez},
  \citenamefont {Aasen},\ and\ \citenamefont {Kempf}}]{AasenPRL}%
  \BibitemOpen
  \bibfield  {author} {\bibinfo {author} {\bibfnamefont {E.}~\bibnamefont
  {Mart\'{i}n-Mart\'{i}nez}}, \bibinfo {author} {\bibfnamefont
  {D.}~\bibnamefont {Aasen}}, \ and\ \bibinfo {author} {\bibfnamefont
  {A.}~\bibnamefont {Kempf}},\ }\href {\doibase 10.1103/PhysRevLett.110.160501}
  {\bibfield  {journal} {\bibinfo  {journal} {Phys. Rev. Lett.}\ }\textbf
  {\bibinfo {volume} {110}},\ \bibinfo {pages} {160501} (\bibinfo {year}
  {2013}{\natexlab{b}})}\BibitemShut {NoStop}%
\bibitem [{\citenamefont {Mart\'{i}n-Mart\'{i}nez}\ \emph
  {et~al.}(2013{\natexlab{c}})\citenamefont {Mart\'{i}n-Mart\'{i}nez},
  \citenamefont {Montero},\ and\ \citenamefont {del Rey}}]{Wavepackets}%
  \BibitemOpen
  \bibfield  {author} {\bibinfo {author} {\bibfnamefont {E.}~\bibnamefont
  {Mart\'{i}n-Mart\'{i}nez}}, \bibinfo {author} {\bibfnamefont
  {M.}~\bibnamefont {Montero}}, \ and\ \bibinfo {author} {\bibfnamefont
  {M.}~\bibnamefont {del Rey}},\ }\href {\doibase 10.1103/PhysRevD.87.064038}
  {\bibfield  {journal} {\bibinfo  {journal} {Phys. Rev. D}\ }\textbf {\bibinfo
  {volume} {87}},\ \bibinfo {pages} {064038} (\bibinfo {year}
  {2013}{\natexlab{c}})}\BibitemShut {NoStop}%
\bibitem [{\citenamefont {Romer}\ and\ \citenamefont {Dicke}(1955)}]{Dicke55}%
  \BibitemOpen
  \bibfield  {author} {\bibinfo {author} {\bibfnamefont {R.~H.}\ \bibnamefont
  {Romer}}\ and\ \bibinfo {author} {\bibfnamefont {R.~H.}\ \bibnamefont
  {Dicke}},\ }\href {\doibase 10.1103/PhysRev.99.532} {\bibfield  {journal}
  {\bibinfo  {journal} {Phys. Rev.}\ }\textbf {\bibinfo {volume} {99}},\
  \bibinfo {pages} {532} (\bibinfo {year} {1955})}\BibitemShut {NoStop}%
\bibitem [{\citenamefont {Briaudeau}\ \emph {et~al.}(1998)\citenamefont
  {Briaudeau}, \citenamefont {Saltiel}, \citenamefont {Nienhuis}, \citenamefont
  {Bloch},\ and\ \citenamefont {Ducloy}}]{Stephen98}%
  \BibitemOpen
  \bibfield  {author} {\bibinfo {author} {\bibfnamefont {S.}~\bibnamefont
  {Briaudeau}}, \bibinfo {author} {\bibfnamefont {S.}~\bibnamefont {Saltiel}},
  \bibinfo {author} {\bibfnamefont {G.}~\bibnamefont {Nienhuis}}, \bibinfo
  {author} {\bibfnamefont {D.}~\bibnamefont {Bloch}}, \ and\ \bibinfo {author}
  {\bibfnamefont {M.}~\bibnamefont {Ducloy}},\ }\href {\doibase
  10.1103/PhysRevA.57.R3169} {\bibfield  {journal} {\bibinfo  {journal} {Phys.
  Rev. A}\ }\textbf {\bibinfo {volume} {57}},\ \bibinfo {pages} {R3169}
  (\bibinfo {year} {1998})}\BibitemShut {NoStop}%
\bibitem [{\citenamefont {Sarkisyan}\ \emph {et~al.}(2004)\citenamefont
  {Sarkisyan}, \citenamefont {Varzhapetyan}, \citenamefont {Sarkisyan},
  \citenamefont {Malakyan}, \citenamefont {Papoyan}, \citenamefont {Lezama},
  \citenamefont {Bloch},\ and\ \citenamefont {Ducloy}}]{Sarkisyan04}%
  \BibitemOpen
  \bibfield  {author} {\bibinfo {author} {\bibfnamefont {D.}~\bibnamefont
  {Sarkisyan}}, \bibinfo {author} {\bibfnamefont {T.}~\bibnamefont
  {Varzhapetyan}}, \bibinfo {author} {\bibfnamefont {A.}~\bibnamefont
  {Sarkisyan}}, \bibinfo {author} {\bibfnamefont {Y.}~\bibnamefont {Malakyan}},
  \bibinfo {author} {\bibfnamefont {A.}~\bibnamefont {Papoyan}}, \bibinfo
  {author} {\bibfnamefont {A.}~\bibnamefont {Lezama}}, \bibinfo {author}
  {\bibfnamefont {D.}~\bibnamefont {Bloch}}, \ and\ \bibinfo {author}
  {\bibfnamefont {M.}~\bibnamefont {Ducloy}},\ }\href {\doibase
  10.1103/PhysRevA.69.065802} {\bibfield  {journal} {\bibinfo  {journal} {Phys.
  Rev. A}\ }\textbf {\bibinfo {volume} {69}},\ \bibinfo {pages} {065802}
  (\bibinfo {year} {2004})}\BibitemShut {NoStop}%
\end{thebibliography}%

\end{document}